\documentclass[10pt,a4paper]{article}
\usepackage[left=20mm, top=20mm, right=20mm, bottom=20mm, nohead, foot=10mm]{geometry}
\usepackage[utf8]{inputenc}
\usepackage[english]{babel}

\usepackage[affil-it]{authblk}
\usepackage{multicol}
\usepackage{caption}

\usepackage[hidelinks]{hyperref}
\usepackage{booktabs}
\usepackage{enumitem}
\usepackage{amsmath}
\usepackage{amssymb}
\usepackage{amsfonts}
\usepackage{mathrsfs}
\usepackage{multicol}
\usepackage{graphicx}
\usepackage{float}
\usepackage{xcolor}

\usepackage[style=numeric,sorting=none,giveninits=true,isbn=false,alldates=year,doi=false,eprint=false]{biblatex}
\renewbibmacro{in:}{}
\AtEveryBibitem{\clearfield{number}\clearfield{issue}}

\usepackage{titlesec}
\titleformat{\section}[block]{\normalfont\sffamily}{}{.5em}{\bfseries}
\titleformat{\subsection}[block]{\normalfont\sffamily}{}{.5em}{\bfseries}

\newenvironment{FigureOneColumn}
  {\par\medskip\noindent\minipage{\linewidth}\captionsetup{type=figure}}
  {\endminipage\par\medskip}

\begin{document}

\title{\sffamily Flux Qubit Based on Hybrid Ferromagnetic-Superconducting Device}

\author[1,2]{\normalsize Filipp N. Rybakov\thanks{philipp.rybakov@physics.uu.se}} 
\author[2]{\normalsize Egor Babaev\thanks{babaev@kth.se}}

\affil[1]{\small
Department of Physics and Astronomy, Uppsala University, SE-75120 Uppsala, Sweden
}

\affil[2]{\small
Department of Physics, KTH Royal Institute of Technology, SE-10691 Stockholm, Sweden
}

\date{}

\clubpenalty=10000
\widowpenalty=10000

\maketitle
\vspace{-3.0\baselineskip}

\renewcommand{\abstractname}{}
\begin{abstract}  
We propose a realization of flux qubit based on the hybrid ferromagnetic-superconducting device where the flux bias is induced purely by vector potential of the vanishing magnetic field. We support our conclusions with theoretical analysis and self-consistent three-dimensional simulations for material specific parameters.
\end{abstract}

\begin{multicols}{2}[]

\section*{Introduction}

One of the promising strategy to realize a quantum bit is based on creation of a superposition of circulating currents in a superconducting device~\cite{doi:10.1126/science.285.5430.1036,PhysRevB.60.15398,Yan2016,PhysRevX.11.021026,somoroff2021millisecond,PhysRevX.11.011010}.
  Such qubit realization, termed ``flux qubit'' is realized by creating a superconducting loop subject to external magnetic field (Fig.~\ref{fig1}a). When the loop encloses (nearly) half of the superconducting flux quantum, ${\Phi\approx\frac{1}{2}\Phi_0}$, there is an energy near degeneracy of clock-wise (CW) and counter-clockwise (CCW) circulation of superconducting currents which is exploited to create a quantum bit~\cite{doi:10.1126/science.285.5430.1036, doi:10.1126/science.1081045, Clarke2008}. Some of the scalability issues are naturally related to necessety of creating stable external field in large area that also should be calibrated.
  Here we propose a realization of flux qubit without the external magnetic field that will provide a compact design.
 
 We propose biasing the device in a superconductor-ferromagnet hybrid realization where the flux bias  is induced purely by vector potential from the  ferromagnet component (Fig.~\ref{fig1}b).
First, we give analytical estimates for the idealized case, and then a self-consistent calculation for realistic material constants and geometries.

\begin{FigureOneColumn}
	\centering
	\includegraphics[width=7.2cm]{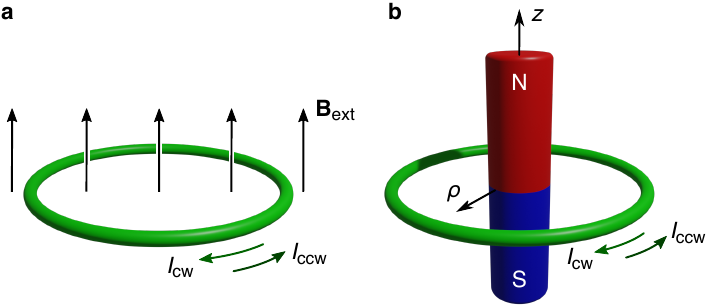}
	\caption{
\textbf{Sketches of flux qubits}.
\textbf{a}, Thin superconducting loop in a uniform external magnetic field. 
\textbf{b}, Same as \textbf{a} but instead of the applied field, the system includes a magnet. 
}
	\label{fig1}
\end{FigureOneColumn}

\begin{figure*}[t]
	\centering
	\includegraphics[width=14.4cm]{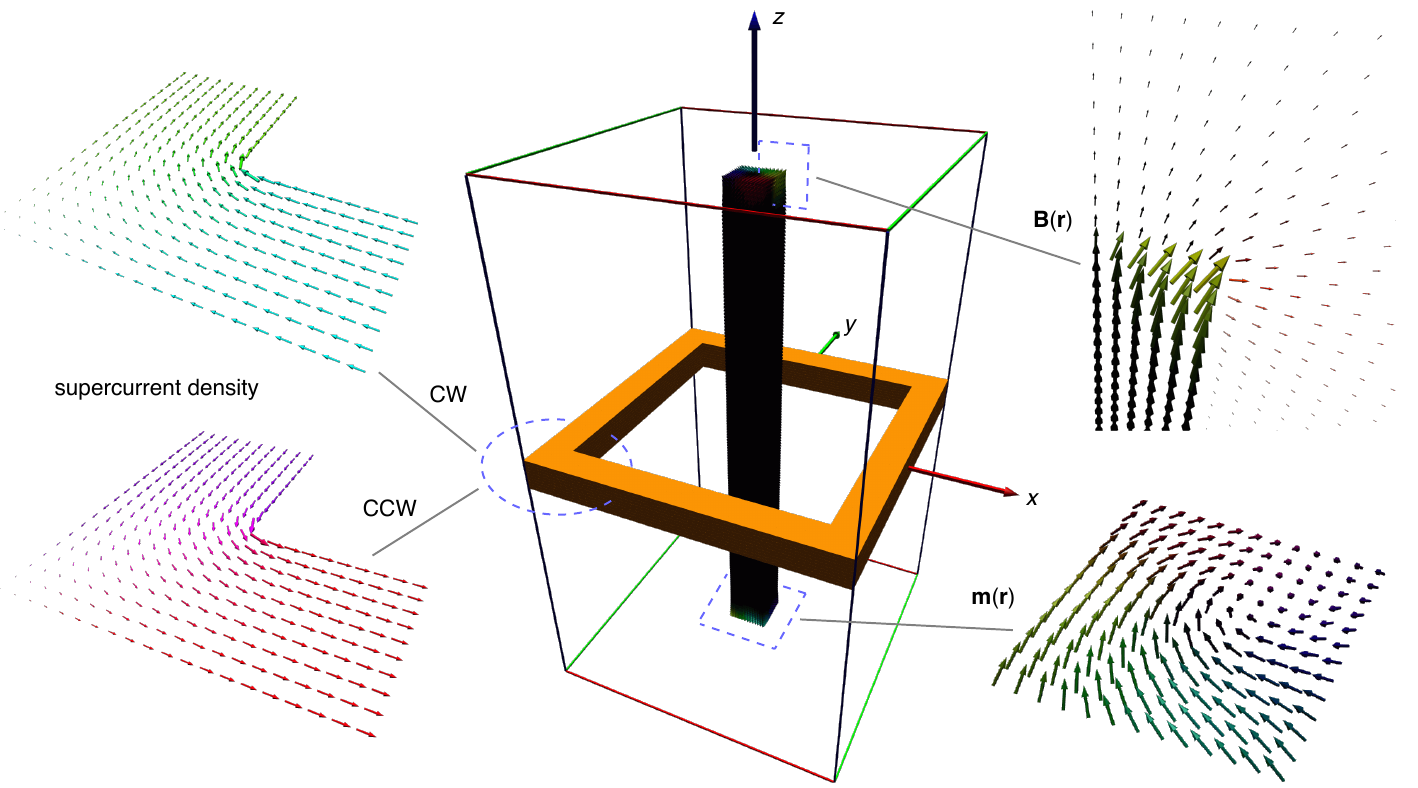}
	\caption{
\textbf{Calculated equilibrium states of a flux qubit based on a ferromagnet}. 
The computing box contains a superconducting domain in the form of a rectangular loop and a centered ferromagnetic cuboid. 
The insets show the distribution of supercurrents near the corner of the loop, as well as the distribution of the magnetic field and magnetization near the ends of the ferromagnet.
}
	\label{fig_simulations}
\end{figure*}

\section*{Theoretical analysis}

In idealized scenarios, i.e. in the case (Fig.~\ref{fig1}a) with a uniform applied magnetic field and in the case (Fig.~\ref{fig1}b) with an infinitely long whisker magnetized to saturation, the vector potential is a function of the distance $\rho$ to the axis of symmetry and, respectively, can be written as
\begin{subequations}
\begin{align}
\mathbf{A} &=  \frac{1}{2} B_\text{ext} \rho \ \hat{\mathbf{e}}_\varphi, \\
\mathbf{A} &=  
\frac{1}{2}{\mu_0 M_\text{s} \rho_\text{m}}  \hat{\mathbf{e}}_\varphi
\begin{cases}
 \frac{\rho}{\rho_\text{m}} \quad (\rho \leq \rho_\text{m}), \\ 
\frac{\rho_\text{m}}{\rho} \quad (\rho_\text{m} < \rho), 
 \end{cases} \label{A_b}
\end{align}
\label{A}
\end{subequations}
 where $\mu_0$ is the vacuum permeability, $M_\text{s}$ is magnetization and $\rho_\text{m}$ is the radius of the whisker. Direct calculation of the magnetic field, ${\mathbf{B} = \mathbf{\nabla}\times\mathbf{A}}$, from~(\ref{A_b}) gives ${B = {\mu_0 M_\text{s}}}$ inside the magnet and ${B = 0}$ in the exterior. 
Despite the zero magnetic fields in the exterior, the vector potential is non-trivial at any distance from the magnet. Accordingly, the condensate of Cooper pairs of electrons in a coaxial superconducting coil is affected by this vector potential. 
The corresponding effect on the condensate energy density is proportional to 
${|\psi|^2|\hbar \mathbf{\nabla}\theta + 2e\mathbf{A}|^2}$, 
where $\theta$ is the phase of the wave function 
${\psi = |\psi|e^{\dot{\iota}\theta}}$  
and $e$ is the absolute electron charge.
Considering that the phase winding, 
${N = \frac{1}{2\pi}\oint \mathbf{\nabla}\theta \cdot d\mathbf{l}}$ 
is an integer, we find that if the setup provides ${A(\rho_\text{s}) = \hbar/4 e \rho_\text{s} }$, then there is a two-degenerate energy minima (corresponding to the values ${N=-1}$ and ${N=0}$). 
In our notation, ${\rho_\text{s}}$ denotes the radius of a negligibly thin superconducting loop.
Taking into account equations~(\ref{A}) and the fact that ${\Phi_0\equiv h/2e\approx 2.07\times10^{-15}}$~Wb, 
the condition of inducing half of a flux quantum by either an externally applied magnetic field or by an enclosed (at infinity) ferromagnet, respectively, is 
\begin{subequations}
\begin{align}
B_\text{ext} \pi \rho_\text{s}^2 & = \frac{1}{2}\Phi_0, \label{hQ_a} \\ 
{\mu_0 M_\text{s}}\pi \rho_\text{m}^2 &= \frac{1}{2}\Phi_0. \label{hQ_b}
\end{align}
\label{hQ}
\end{subequations}
It is important to emphasize that among the geometrical parameters, the radius of the superconducting loop plays a fundamental role in the first case, while in the second case it is the radius of the magnetic whisker.
Accordingly, in the first case, taking into account~(\ref{hQ_a}), the compact design entails an undesirable strong increase in the bias field. 
While in the case of using a ferromagnet, taking into account~(\ref{hQ_b}), the radius of the superconducting loop can theoretically be reduced down to tens of nanometers, due to the fact that the field inside the magnet, ${\mu_0 M_\text{s}}$ can be about~2~Tesla.

The above argument is based on simple generic considerations. 
However, to implement the device in practice, one needs to address several problems which are not present in the case of bias magnetic fields. 
Namely, because the device does not involve external tuning tools, it must be designed as accurately as possible in relation to the main requirement -- the energy (near-) equivalence of two quantum states.
Accordingly, it is important to take into account a number of features that matter when considering realistic materials and shapes. For example, a magnetic whisker must have a finite size. Which, in turn, entails stray fields, and also the magnetization inside the whisker becomes inhomogeneous. In addition, the supercurrent induces its own magnetic field, which contributes to the energy balance. Finally, for circuit elements, rectangular shapes may be preferable to cylindrical ones. 
Below, we demonstrate that these issues are addressed using self-consistent material-specific simulation of the device.
We have designed an example device in such a way as to achieve matching the energies of quantum states with clock-wise and counter-clock-wise currents.

\section*{Simulations}

The static Hamiltonian for a hybrid system of superconductors and magnets coupled through magnetic potential could be written as:
\begin{align}
H\!=\!& \int\limits_{\Omega_\text{s}} d\mathbf{r}  
\Bigg( \frac{1}{2m_{*}}| \left( -\dot{\iota} \hbar \mathbf{\nabla} - q_{*} {\mathbf A}   \right)\psi|^2 + \alpha |\psi|^2 + \frac{\beta}{2}|\psi|^4 \Bigg)
\nonumber\\ 
& + \int\limits_{\Omega_\text{m}} d\mathbf{r}  
\Bigg( \mathcal{A}\,\sum_{i=x,y,z}|\nabla m_i|^2 - M_\text{s}\  \mathbf{m}\cdot({\mathbf \nabla}\times{\mathbf A}) \Bigg)
\nonumber\\ 
 & + \frac{1}{2\mu_0}\int\limits_{\mathbb{R}^3} d\mathbf{r}  
\sum_{i=x,y,z}|\nabla A_i|^2, 
\label{H}
\end{align}
where $\Omega_\text{s}$ and $\Omega_\text{m}$ -- superconducting and magnetic domains respectively, and $\mathbb{R}^3$ denotes three-dimensional space. 
Expression~(\ref{H}) is a combination of the Ginzburg-Landau and micromagnetic functionals based on the uniform consideration of the vector potential~\cite{doi:10.1137/S0036142997330317,doi:10.1137/19M1261365}.
The parameter $m_{*}$ denotes the effective mass of the Cooper pair whose effective charge is ${q_{*}=-2e}$.
Phenomenological parameters $\alpha<0$ and $\beta$ are representable via the critical field, $B_\text{c}$ and the Ginzburg-Landau parameter, $\kappa$: 
${\alpha=-\frac{B_\text{c}\kappa\hbar |q_{*}|}{\sqrt{2}m_{*}}}$, 
${\beta=\frac{\kappa^2\hbar^2 \mu_0 q_{*}^2}{2m_{*}^2}}$.
The parameter $\mathcal{A}$ denotes the exchange stiffness of a ferromagnet.
The unit vector field, $\mathbf{m}(\mathbf{r})$ represents the magnetization texture in a ferromagnet.

We consider (as $\Omega_\text{s}$) a superconducting rectangular loop  200~nm wide with squared 20$\times$20~nm cross-section. 
We choose the material parameters of the superconductor loop similar to those fitted for nanosamples of aluminum at low temperatures~\cite{vanWoerkom2015}: ${B_\text{c}=0.02}$~T and  ${\kappa=1}$. 
It is important to note that the Ginzburg-Landau functional with appropriately fitted parameters can provide quantitative description of a superconductor whose temperature is much lower than the critical one~\cite{PhysRevLett.93.257002, PhysRevB.85.134514}.

We consider (as $\Omega_\text{m}$) a centered ferromagnetic whisker of size {24$\times$24$\times$(280 -- 330)}~nm. 
Similar Fe- and Co-based nano-whiskers can be synthesized, for example, via physical vapor deposition~\cite{HUANG2021109914}.
We choose the material parameters of the ferromagnet similar to those for iron~\cite{6497624}: ${\mathcal{A}=10^{-11}}$~J~m$^{-1}$, ${M_\text{s} = 1.7\times10^6}$~A~m$^{-1}$.

The solutions for~(\ref{H}) were obtained by numerical minimization technique by using Excalibur software~\cite{excalibur}. A summary of the underlying discretization and minimization methods can be found in~\cite{Rybakov_thesis}. 
The discretization grid parameter was equal to 2~nm in all three dimensions. 
The expression~(\ref{H}) has been converted to a dimensionless form so that $\psi$ is assumed to be in the units of $\sqrt{|\alpha|/\beta}$, distances -- in the units of $l_0=10^{-9}$~m, vector potential -- in the units of ${B_0 l_0}$, energy density -- in the units of $B_0^2/\mu_0$, where $B_0=1$~T.

By adjusting the length of the magnetic whisker, we found that the length corresponding to the equality of the total energies for CW and CCW currents is approximately equal to 316~nm. 
Figure~\ref{fig_simulations} illustrates the corresponding numerical solution. 
We note that calculations show that the supercurrent density is almost uniformly distributed over the cross-section of the superconductor (only slightly except at the corners).

Our fully-self-consistent calculations confirm that this ferromagnetic material of this size and shape is in a single-domain state, hence allowing this qubit set up. 
Furthermore, as shown in the insets of the Figure~\ref{fig_simulations}, the stray fields are relatively small. At the same time, the magnetization near the ends is substantially inhomogeneous. 
However, we find that  the stability of the magnetic state is very high, and the initial guess with a two-domain configuration and a domain wall in the middle of the upper/lower half of the whisker is unstable and minimizes to the solution shown in Figure~\ref{fig_simulations}.

\section*{Conclusion}

In conclusion, by using a fully self-consistent simulation of superconducting loop coupled to a ferromagnet we demonstrated that it is possible to realize a quantum bit where the flux is induced by a ferromagnet. 
The compact design, due to high internal magnetic fields in the ferromagnet, and at the same time the robustness of this magnetic state is a potential advantage of such a device, and the degree of freedom associated with the almost arbitrary loop area opens up a wide range for circuit engineering.
The shown numerical solvability of the problem also allows designing more complicated geometries such as linked superconducting and ferromagnetic loops that would diminish stray fields of such a qubit.

\section*{Acknowledgments}

The authors acknowledge support from the Swedish Research Council (including grants No. 2016-06122, 2018-03659), Olle Engkvists stiftelse and VINNOVA.

\printbibliography

@article{PhysRevLett.93.257002,
  title = {Experimental Evidence for Giant Vortex States in a Mesoscopic Superconducting Disk},
  author = {Kanda, A. and Baelus, B. J. and Peeters, F. M. and Kadowaki, K. and Ootuka, Y.},
  journal = {Phys. Rev. Lett.},
  volume = {93},
  issue = {25},
  pages = {257002},
  numpages = {4},
  year = {2004},
  month = {Dec},
  publisher = {American Physical Society},
  doi = {10.1103/PhysRevLett.93.257002},
  url = {https://link.aps.org/doi/10.1103/PhysRevLett.93.257002}
}

@article{PhysRevB.85.134514,
  title = {Microscopic derivation of two-component Ginzburg-Landau model and conditions of its applicability in two-band systems},
  author = {Silaev, Mihail and Babaev, Egor},
  journal = {Phys. Rev. B},
  volume = {85},
  issue = {13},
  pages = {134514},
  numpages = {8},
  year = {2012},
  month = {Apr},
  publisher = {American Physical Society},
  doi = {10.1103/PhysRevB.85.134514},
  url = {https://link.aps.org/doi/10.1103/PhysRevB.85.134514}
}

@phdthesis{Rybakov_thesis,
  author       = {Rybakov, Filipp N.}, 
  title        = {Topological excitations in field theory models of superconductivity and magnetism},
  school       = {KTH Royal Institute of Technology},
  year         = 2021,
  address      = {Stockholm},
  url = {http://urn.kb.se/resolve?urn=urn:nbn:se:kth:diva-301652}
}

@misc{excalibur,
author = {F. N. Rybakov and E. Babaev},
title = {Excalibur software},
note = {\\URL: \href{http://quantumandclassical.com/excalibur/}{http://quantumandclassical.com/excalibur/}}
}

@Article{Yan2016,
author={Yan, Fei
and Gustavsson, Simon
and Kamal, Archana
and Birenbaum, Jeffrey
and Sears, Adam P.
and Hover, David
and Gudmundsen, Ted J.
and Rosenberg, Danna
and Samach, Gabriel
and Weber, S.
and Yoder, Jonilyn L.
and Orlando, Terry P.
and Clarke, John
and Kerman, Andrew J.
and Oliver, William D.},
title={The flux qubit revisited to enhance coherence and reproducibility},
journal={Nature Communications},
year={2016},
month={Nov},
day={03},
volume={7},
number={1},
pages={12964},
issn={2041-1723},
doi={10.1038/ncomms12964},
url={https://doi.org/10.1038/ncomms12964}
}

@article{
doi:10.1126/science.285.5430.1036,
author = {J. E. Mooij  and T. P. Orlando  and L. Levitov  and Lin Tian  and Caspar H. van der Wal  and Seth Lloyd },
title = {Josephson Persistent-Current Qubit},
journal = {Science},
volume = {285},
number = {5430},
pages = {1036-1039},
year = {1999},
doi = {10.1126/science.285.5430.1036},
URL = {https://www.science.org/doi/abs/10.1126/science.285.5430.1036},
}

@article{PhysRevX.11.011010,
  title = {Universal Fast-Flux Control of a Coherent, Low-Frequency Qubit},
  author = {Zhang, Helin and Chakram, Srivatsan and Roy, Tanay and Earnest, Nathan and Lu, Yao and Huang, Ziwen and Weiss, D. K. and Koch, Jens and Schuster, David I.},
  journal = {Phys. Rev. X},
  volume = {11},
  issue = {1},
  pages = {011010},
  numpages = {13},
  year = {2021},
  month = {Jan},
  publisher = {American Physical Society},
  doi = {10.1103/PhysRevX.11.011010},
  url = {https://link.aps.org/doi/10.1103/PhysRevX.11.011010}
}

@article{somoroff2021millisecond,
      title={Millisecond coherence in a superconducting qubit}, 
      author={Aaron Somoroff and Quentin Ficheux and Raymond A. Mencia and Haonan Xiong and Roman V. Kuzmin and Vladimir E. Manucharyan},
    journal = {arXiv:2103.08578},
    year = {2021},
  url = {https://arxiv.org/abs/2103.08578}
}

@article{PhysRevX.11.021026,
  title = {Fast Logic with Slow Qubits: Microwave-Activated Controlled-Z Gate on Low-Frequency Fluxoniums},
  author = {Ficheux, Quentin and Nguyen, Long B. and Somoroff, Aaron and Xiong, Haonan and Nesterov, Konstantin N. and Vavilov, Maxim G. and Manucharyan, Vladimir E.},
  journal = {Phys. Rev. X},
  volume = {11},
  issue = {2},
  pages = {021026},
  numpages = {16},
  year = {2021},
  month = {May},
  publisher = {American Physical Society},
  doi = {10.1103/PhysRevX.11.021026},
  url = {https://link.aps.org/doi/10.1103/PhysRevX.11.021026}
}

@article{PhysRevB.60.15398,
  title = {Superconducting persistent-current qubit},
  author = {Orlando, T. P. and Mooij, J. E. and Tian, Lin and van der Wal, Caspar H. and Levitov, L. S. and Lloyd, Seth and Mazo, J. J.},
  journal = {Phys. Rev. B},
  volume = {60},
  issue = {22},
  pages = {15398--15413},
  numpages = {0},
  year = {1999},
  month = {Dec},
  publisher = {American Physical Society},
  doi = {10.1103/PhysRevB.60.15398},
  url = {https://link.aps.org/doi/10.1103/PhysRevB.60.15398}
}

@article{doi:10.1126/science.1081045,
author = {I. Chiorescu  and Y. Nakamura  and C. J. P. M. Harmans  and J. E. Mooij },
title = {Coherent Quantum Dynamics of a Superconducting Flux Qubit},
journal = {Science},
volume = {299},
number = {5614},
pages = {1869-1871},
year = {2003},
doi = {10.1126/science.1081045},
URL = {https://www.science.org/doi/abs/10.1126/science.1081045},
eprint = {https://www.science.org/doi/pdf/10.1126/science.1081045}
}

@article{Clarke2008,
author={Clarke, John
and Wilhelm, Frank K.},
title={Superconducting quantum bits},
journal={Nature},
year={2008},
month={Jun},
day={01},
volume={453},
number={7198},
pages={1031-1042},
issn={1476-4687},
doi={10.1038/nature07128},
url={https://doi.org/10.1038/nature07128}
}

@article{HUANG2021109914,
title = {Synthesis of magnetic Fe and Co nano-whiskers and platelets via physical vapor deposition},
journal = {Materials and Design},
volume = {208},
pages = {109914},
year = {2021},
issn = {0264-1275},
doi = {https://doi.org/10.1016/j.matdes.2021.109914},
url = {https://www.sciencedirect.com/science/article/pii/S0264127521004676},
author = {Wenting Huang and Christophe Gatel and Zi-An Li and Gunther Richter}
}

@article{doi:10.1137/S0036142997330317,
author = {Du, Qiang and Wu, Xiaonan},
title = {Numerical Solution of the Three-Dimensional Ginzburg--Landau Models Using Artificial Boundary},
journal = {SIAM Journal on Numerical Analysis},
volume = {36},
number = {5},
pages = {1482-1506},
year = {1999},
doi = {10.1137/S0036142997330317},
URL = {https://doi.org/10.1137/S0036142997330317},
eprint = {https://doi.org/10.1137/S0036142997330317}
}

@article{doi:10.1137/19M1261365,
author = {Di Fratta, Giovanni and Muratov, Cyrill B. and Rybakov, Filipp N. and Slastikov, Valeriy V.},
title = {Variational Principles of Micromagnetics Revisited},
journal = {SIAM Journal on Mathematical Analysis},
volume = {52},
number = {4},
pages = {3580-3599},
year = {2020},
doi = {10.1137/19M1261365},
URL = {https://doi.org/10.1137/19M1261365},
eprint = {https://doi.org/10.1137/19M1261365}
}

@article{6497624,
  author={Abo, Gavin S. and Hong, Yang-Ki and Park, Jihoon and Lee, Jaejin and Lee, Woncheol and Choi, Byoung-Chul},
  journal={IEEE Transactions on Magnetics}, 
  title={Definition of Magnetic Exchange Length}, 
  year={2013},
  volume={49},
  number={8},
  pages={4937-4939},
  doi={10.1109/TMAG.2013.2258028}
}

@article{vanWoerkom2015,
author={van Woerkom, David J.
and Geresdi, Attila
and Kouwenhoven, Leo P.},
title={One minute parity lifetime of a NbTiN Cooper-pair transistor},
journal={Nature Physics},
year={2015},
month={Jul},
day={01},
volume={11},
number={7},
pages={547-550},
issn={1745-2481},
doi={10.1038/nphys3342},
url={https://doi.org/10.1038/nphys3342}
}

\end{multicols}

\end{document}